\documentclass[11pt,a4paper]{article}
\usepackage[english]{babel}
\usepackage{amsmath,amsthm,amssymb,epsfig,latexsym}
\usepackage[numbers,square,sort&compress]{natbib}
\usepackage{color}

\newcommand{\be}[1]{\begin{equation}\label{#1}}
\newcommand{\ba}[1]{\begin{multline}\label{#1}}
\newcommand{\ee}{\end{equation}}
\newcommand{\ea}{\end{eqnarray}}

 \makeatletter
 \@addtoreset{equation}{section}
 \makeatother
 
\begin{document}

\vspace{12pt}

\begin{center}
\begin{LARGE}
{\bf Comments to the \\[1.ex]
"Theory of Thomson scattering in inhomogeneous media" }
\end{LARGE}

\vspace{40pt}

\begin{large}
{V.~V.~Belyi${}^{a}$\footnote{sbelyi@izmiran.ru}}
\end{large}

 \vspace{12mm}

\vspace{4mm}

${}^a$ {\it IZMIRAN, Russian Academy of Sciences, Troitsk, Moscow, 108840, Russia}

\vspace{4mm}

\end{center}


\vspace{4mm}


\begin{abstract}
In this short communication, we draw the readers' attention to the inconsistency in the derivation of the Thomson scattering spectrum in inhomogeneous plasma, which leads to violation of the Fluctuation-Dissipation Theorem and a substantial deviation from the results of the rigorous kinetic theory. Moreover, the self-consistent kinetic theory predicts the asymmetry of the spectral lines in inhomogeneous plasma.
\end{abstract}

\vspace{1cm}

\vspace{2mm}

\section{Introduction}
On 12 April 2016 the Journal of Scientific Reports published the
 article: P.M. Kozlowski, B.J.B. Crowley, D.O. Gericke, S.P. Regan
 and G. Gregori, "Theory of Thomson scattering in inhomogeneous media" \cite{Kozlowski}. 
The main result of paper has been based on the Eq.:
\begin{equation}
S({\bf k,}\omega )=\frac{S({\bf k,}\omega )^{id}}{\left\vert \epsilon ({\bf %
k,}\omega )\right\vert ^{2}},  \label{a.3}
\end{equation}%
"where $S({\bf k,}\omega )$ is the total dynamic structure factor, $S({\bf k,}%
\omega )^{id}$ is the dynamic structure factor for an ideal (noniteracting
gas), and the dielectric (screening) function $\epsilon ({\bf %
k,}\omega )$ in the denominator of Eq. (\ref{a.3}) in the first order gradient expansion over slow time and space is:
\begin{equation}
\epsilon ({\bf k,}\omega )=1+\chi({\bf k,}\omega)=1+(1+i\frac{\partial }{\partial \omega }\frac{%
\partial }{\partial \mu t}-i\frac{\partial }{\partial \mu {\bf r}}\cdot 
\frac{\partial }{\partial {\bf k}})\chi ^{eq}({\bf k,}\omega,\mu t,\mu {\bf %
r}),  \label{a.7}
\end{equation}%
 $%
\chi ^{eq}$ is the susceptibility of the ideal Coulomb plasma. 
The index "eq" labels the susceptibility for a homogeneous system in thermodynamic equilibrium"\cite{Kozlowski}.
Herewith the structure factor $S({\bf k,}\omega )^{id}$ in Eq. (\ref{a.3}) remains unchanged ($sic!$).
 Regretfully, instead of constructing a self-consistent kinetic theory
 of Thomson scattering in a non-uniform plasma, the authors, on an
 ad hoc basis, expanded in a small
 parameter of inhomogeneity and nonstationarity the dielectric
 permeability in the denominator of the well-known expression for the
 spectral function of a stationary and uniform plasma \cite{Ishimaru}. This caused two 
consequences: First, the obtained result is nonphysical,
 since it contradicts the basic physical principle -- the
 Fluctuation-Dissipation Theorem (FDT) in the local equilibrium state. 
FDT for a local equilibrium state was proved by
Balescu \cite{Balescu}. In the local equilibrium state the parameters 
of a system can be changed adiabatically
 on a scale greater than the particle mean free path. Inhomogeneity and 
nonstationarity in the plasma fluctuations is manifested via a non-local
 dependence upon coordinates and time (Non-Markov) \cite{BKW}. 
FDT for a non-local plasma was given in our paper \cite{Belyi}.  
Second, the obtained correction due
 to the inhomogeneity is erroneous for the Langmuir oscillations, especially
 for small wave numbers $k<k_{D}$, which usually occurs in experiments.
And last but not least: the rigorous kinetic theory predicts \cite{Belyi} 
the asymmetry of the spectral lines in inhomogeneous plasma.
Such asymmetry has been indeed detected in experiments \cite{Strun, BelyiStrun} ,
 but does not appear in the incorrect approach of \cite{Kozlowski}.

\section*{Results}

For a correct description of non-local effects in inhomogeneous plasma the
kinetic approach should be applied. Indeed, using Klimontovich-Langevin method \cite{KLim},
for describing kinetic fluctuation, as well as a multiple space and time scale 
analysis we have shown \cite{Belyi} that the spectral function of electrostatic
 field fluctuations $(\delta {\bf E}%
\delta {\bf E})_{\omega ,{\bf k}}$ for the local equilibrium state is determined
by the following expression:  
\begin{equation}
(\delta {\bf E}\delta {\bf E})_{\omega ,{\bf k}}=\sum\limits_{a}\frac{8\pi \ \Theta_{a}
{\rm {Im}}\,\chi_{a} ({\bf k,}\omega)}{\omega_{a} \left\vert \epsilon ({\bf k,}\omega
)\right\vert ^{2}{\bf \ }}.  \label{a.8}
\end{equation}
Here
\begin{equation}
\epsilon ({\bf k,}\omega )=1+\sum\limits_{a}\chi_{a}({\bf k,}\omega);\text{ \ \ }
\end{equation}%
\begin{equation}
\chi_{a}({\bf k,}\omega)=(1+i\mu\frac{\partial }{\partial \omega }\frac{%
\partial }{\partial \mu t}-i\mu\frac{\partial }{\partial \mu {\bf r}}\cdot 
\frac{\partial }{\partial {\bf k}})\chi_{a} ^{leq}({\bf k,}\omega,\mu t,\mu {\bf %
r}),       
\end{equation}%
$\chi_{a} ^{leq}$ - is the local equilibrium susceptibility. 
$\omega_{a}=\omega-{\bf k}{\bf V_{a}}$,  $\Theta_{a} $ is 
the temperature in energy units. Eq. (\ref{a.8}) satisfies the FDT, 
when ${\bf V_{a}}=0$ and $\Theta_{a}=\Theta $. Imaginary part of the 
susceptibility $\chi({\bf k,}\omega)$
 determines the width of the spectral line $(\delta {\bf E}\delta {\bf E})_{\omega ,{\bf k}}$
 near the resonance:
\begin{equation}
\gamma =(Im\chi ^{leq}+\mu\frac{\partial }{\partial \omega }\frac{\partial }{%
\partial \mu t}Re\chi ^{leq}-\mu\frac{\partial }{\partial \mu \mathbf{r}}\cdot 
\frac{\partial }{\partial \mathbf{k}}Re\chi ^{leq})/\frac{\partial }{\partial
\omega }Re\chi ^{leq}. \label{a.8a}
\end{equation}%
In Eq. (\ref{a.8a}) there appear 
additional terms at first order of the small parameter $\mu. $
It is important to note that the $imaginary$ part of the dielectric 
susceptibility is now replaced by the $real$ part, which in the plasma resonance may be greater than 
the $imaginary$ part by the same factor $ \mu^{-1}$. Therefore, the second and third terms in Eq. (\ref{a.8a})
in the kinetic regime have an effect comparable to that of the first term. At second order in the expansion
 in $\mu $ the corrections appear only in the $imaginary$ part of the susceptibility, and they can reasonably be neglected. The width of the 
spectral lines Eq. (\ref{a.8a}) is affected by new nonlocal terms. 
They are not related to Joule dissipation and appear because of an additional phase
 shift between the vector of induction and the electric field. This phase shift results 
from the finite time needed to set the polarization in the plasma with dispersion. 
Such a phase shift in the plasma with space dispersion appears due to the medium 
inhomogeneity.
 For the case where the system parameters are
homogeneous in space but vary in time, the correction to the width of the spectral
lines in Eq. (\ref{a.8a}) is
still symmetric with respect to the change of sign of $\omega.$ 
However, when the plasma parameters are space dependent this symmetry is lost. 
The real part of the susceptibility $\chi ^{leq}({\bf k,}\omega)$ in Eq. (\ref{a.8a}) 
is an even function of $\omega. $ This property implies that the contribution of the 
space derivative to the expression for the width
 of the spectral lines is odd function of $\omega.$ Besides this term gives rise 
to an anisotropy in $\bf k $ space.

For the spatially homogeneous case there is no difference between the
spectral properties of the longitudinal electric field and of the electron
density, because they are related by the Poisson equation. This statement is no
longer valid when considering an inhomogeneous plasma. Indeed the
longitudinal electric field is linked to the particle density by the
nonlocal relation:
\begin{equation}
\delta {\bf E}({\bf r},t)=-\frac{\partial }{\partial {\bf r}}%
\sum\limits_{a}e_{a}\int \frac{1}{\left\vert {\bf r-r}^{\prime }\right\vert }%
\delta n_{a}({\bf r}^{\prime },t)d{\bf r}^{\prime }.  \label{a.9}
\end{equation}
In the same approximation as in Eq. (\ref{a.8}) the expression for 
the electron structure factor for a two-component ($a=e,i$) plasma has 
the form \cite{Belyi}:
\begin{equation}
S^{e}({\bf k,}\omega )=\frac{2n_{e}k^{2}}{\omega_{e} k_{D}^{2}}\left\vert \frac{%
1+\widetilde{\chi }_{i}({\bf k,}\omega )}{\widetilde{\epsilon }({\bf k,}%
\omega )}\right\vert ^{2}{\rm {Im}}\,\widetilde{\chi }_{e}({\bf k,}\omega )+\left\vert 
\frac{\widetilde{\chi }_{e}({\bf k,}\omega )}{\widetilde{\epsilon }({\bf k,}%
\omega )}\right\vert ^{2}\frac{\Theta _{i}}{\Theta _{e}}\frac{2n_{e}k^{2}}{%
\omega_{i} k_{D}^{2}}{\rm {Im}}\,\widetilde{\chi }_{i}({\bf k,}\omega ), \label{a.10}
\end{equation}
where $k_{D}$ is the inverse Debye lenth,
\begin{equation}
\widetilde{\epsilon }({\bf k,}\omega )=1+\sum\limits_{a}\widetilde{\chi }%
_{a}({\bf k,}\omega ),  \label{a.11}
\end{equation}
\begin{equation}
\widetilde{\chi }_{a}(\omega ,{\bf k})=(1+i\mu\frac{\partial }{\partial \omega }%
\frac{\partial }{\partial \mu t}-i\mu\frac{1}{k^{2}}\frac{\partial }{\partial
\mu r_{i}}k_{j}\frac{\partial }{\partial k_{i}}k_{j})\chi _{a}^{leq}(\omega ,%
{\bf k},\mu t,\mu {\bf r}).  \label{a.12}
\end{equation}
The inhomogeneous correction in Eq. (\ref{a.12}) $ (\frac{1}{k^{2}}\frac{\partial }{\partial
\mu r_{i}}k_{j}\frac{\partial }{\partial k_{i}}k_{j}Re\chi _{a}^{leq})$ 
is not the same as in Eq. (%
\ref{a.8a}) $(\frac{\partial }{\partial \mu {\bf r}}\cdot 
\frac{\partial }{\partial {\bf k}}Re\chi^{leq})$ and for the plasma mode 
($\omega=\omega_{L} $) \ is greater than the one in Eq. (\ref{a.8a}), particularly 
for $k<k_{D}$, which usually occurs in experiment, by the factor 
$3/2(1+$ $k_{D}^{2}/9k^{2}).$\ 
The origin of this difference is that the Green functions for electrostatic field 
fluctuation and density particle fluctuations are not the same in inhomogeneous
situation. We see (Fig 1) that the asymmetry of the spectral lines is present
both for  $S^{e}({\bf k,}\omega )$ and for $(\delta {\bf E}\delta {\bf E}%
)_{\omega ,{\bf k}}.$ However, this effect is more pronounced in $S^{e}({\bf %
k,}\omega )$ than in $(\delta {\bf E}\delta {\bf E})_{\omega ,{\bf k}}$. Such asymmetry has
been detected in inhomogeneous plasma \cite{Strun,BelyiStrun}.
The asymmetry of lines $S^{e}({\bf k,}\omega )$ can   used as a new diagnostic tool 
to measure local gradients in the plasma by Thomson scattering.

\section*{Acknowledgements}
Fruitful discussions with N. Brilliantov are gratefully acknowledged. Thanks to T. Beuermann for detecting a misprint in \cite{Belyi}.

\begin{figure}[hb]
\begin{center}
\includegraphics[width=0.5\textwidth]{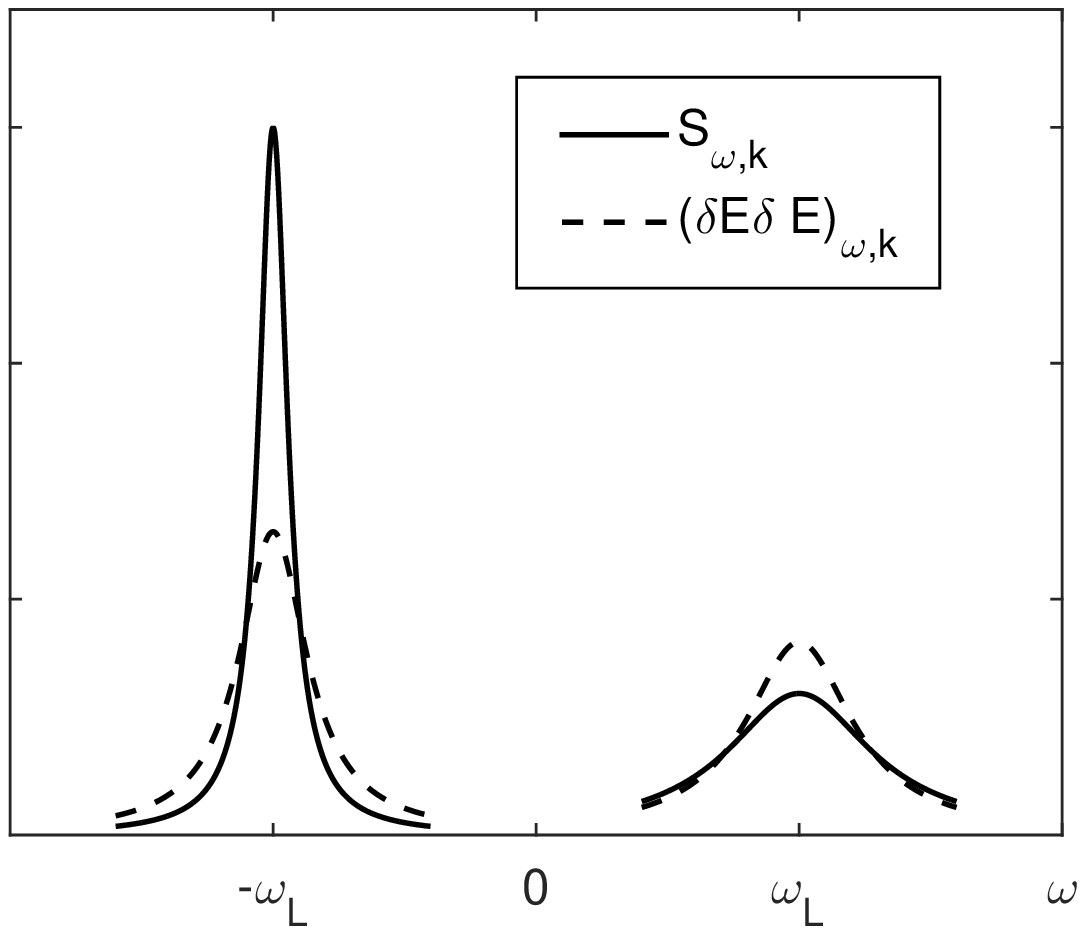}
\caption{The electron structure factor $S^{e}({\bf k,}\omega )$ ( solid line) and the spectral function of electrostatic field
fluctuations $(\delta {\bf E}\delta {\bf E})_{\omega ,{\bf k}}$ (dashed
line) as a function of frequency. $\frac{k_{D}}{k}=3$; ${\bf k\cdot }\frac{\partial n}{\partial
{\bf r}}=\frac{\nu _{ei}nk_{D}^{2}}{27\omega _{L}}$ (The Knudsen number Kn=1/9).
}
\end{center}
\end{figure}
\end{document}